\documentclass[10pt,twocolumn,aps,pra,amsmath,amssymb,showpacs]{revtex4-2}
\usepackage{bm}
\usepackage{mathrsfs}
\usepackage{graphicx}
\usepackage[usenames,dvipsnames,svgnames,table]{xcolor}
\usepackage[unicode=true,
            pdfusetitle, 
            bookmarks=true,
            bookmarksnumbered=false,
            bookmarksopen=false,
            breaklinks=true,
            pdfborder={0 0 0},
            backref=false,
            colorlinks=true]{hyperref}
\hypersetup{linkcolor=NavyBlue,urlcolor=NavyBlue,citecolor=NavyBlue}

\usepackage{amsthm}

\providecommand{\propositionname}{Proposition}

\usepackage{tikz-cd}

\newcommand{\cH}{\mathcal{H}}

\begin{document}

\title{Entanglement recovering  under local  non-positive  reduced dynamics}
\author{Iman Sargolzahi}
\email{sargolzahi@um.ac.ir}
\affiliation{Department of Physics, Faculty of Science, Ferdowsi University of Mashhad, Mashhad, Iran}

\begin{abstract}
 Consider a bipartite quantum system $S=AB$ such that each part interacts only with its local  environment. Under such circumstances, one expects that the entanglement between parts $A$ and $B$ does not increase after the time evolution. In fact, this is the case if the reduced dynamics of the system is given by $\mathcal{E}_{A}\otimes \mathcal{E}_{B}$, where  $\mathcal{E}_{A}$ and $\mathcal{E}_{B}$ are quantum channels, i.e., completely positive trace-preserving maps.
  But, the reduced dynamics of the system may be given by a map as $\Psi_{A}\otimes \Psi_{B}$, where  $\Psi_{A}$ and $\Psi_{B}$ are local non-positive maps. Then, the entanglement between  $A$ and $B$ can  exceed its initial value, as was shown in the case studied by Jordan  \textit{et al.} [ \href{https://journals.aps.org/pra/abstract/10.1103/PhysRevA.76.022102} {Phys. Rev. A \textbf{76}, 022102 (2007)}].
 In this paper, we first explore the general circumstances under which one can find such cases as they found. Next, we introduce another general procedure  which leads to local non-positive maps that exhibit entanglement exceeding.
 Furthermore, we demonstrate that this apparent entanglement increase in the two above procedures  corresponds physically to the retrieval of pre-existing inaccessible entanglement stored
across the global system-environment degrees of freedom, rather than genuine
generation of the  entanglement \textit{ex nihilo}.
 \end{abstract}


\maketitle

\section{Introduction}
Consider a  bipartite quantum system $S=AB$ such that the parts $A$ and $B$ do not interact with each other, but each part interacts with its local  environment $E_A$ and $E_B$, respectively.
In such open quantum systems, entanglement between parts $A$ and $B$ can reveal interesting features such as entanglement  sudden death and entanglement revival  \cite{1, 2, 3, 4, 5, 6}.

Another interesting feature in such systems is that 
entanglement  can even exceed its initial value. Different causes can lead to entanglement exceeding which we will review in the following.

 First is the case studied in Ref. \cite{7}, in which the whole system $S=AB$ evolves under the Hamiltonian $H_{AB}=H_{A} \otimes I_{B}$, where $H_{A}$ is a parity-time symmetric Hamiltonian on $\cH_{A}$, which 
 is a  non-Hermitian operator, and $I_{B}$ is the identity operator on $\cH_{B}$. ($\cH_{A}$ and  $\cH_{B}$ are the Hilbert spaces of the subsystems $A$ and $B$, respectively.) Then, the non-Hermicity of $H_{A}$ leads to entanglement exceeding than its initial value. In fact,  under the time evolution generated by  $H_{AB}$, the reduced state of the part $B$ does not remain unchanged  \cite{7}. 
Note that, since $H_{A}$ is  non-Hermitian, the evolution generated by it is not trace-preserving on part $A$. This results in changing the reduced state of the part $B$, even though the evolution is local. In other words, a non-trace-preserving local evolution has non-local effects which can lead to entanglement increasing.

Second is the case studied in, e.g., Ref. \cite{8}. During the derivation of a master equation for an open quantum system interacting with its environment one usually uses some approximations \cite{9, 10}. These approximations may lead to a map which is not positive \cite{8}, i.e., it does not map all initial states  to valid final states. To solve this problem, one restricts the use of such   map to those initial states (initial density operators) which are mapped to density operators. In Ref. \cite{8}, it has been shown that even restricting the set of possible initial states to this seemingly admissible set may lead to inconsistencies.

In the case studied in Ref. \cite{8}, the system $S$ includes two qubits $A$ and $B$, where $A$ interacts with its environment $E_A$, and $B$ evolves trivially. So, the whole evolution is as $U_{AE_A} \otimes I_B$, where $U_{AE_A}$ is a unitary operator on $\cH_{A}\otimes \cH_{E_A}$. ($ \cH_{E_A}$ is the  Hilbert space of $E_A$.) In addition, the initial state of the whole system-environment is as $\rho_{AB}\otimes \tilde{\omega}_{E_A}$, where $\rho_{AB}$ and $\tilde{\omega}_{E_A}$ are states on $\cH_{A}\otimes\cH_{B}$ and $ \cH_{E_A}$, respectively. Under such circumstances, entanglement between qubits $A$ and $B$ cannot exceed its initial value. 
But, if we use the approximate non-positive map for the reduced dynamics of the system $S=AB$ (for an initial state $\rho_{AB}$ which is mapped to a density operator by this map), this can lead to exceeding the entanglement between  $A$ and $B$, which is a wrong result \cite{8}.

In the above case, entanglement exceeding is a sign of a problem. Also, the first case above is not in the context of  the conventional quantum mechanics, and leads to non-local effects from local interactions. In the following, we will mention two other cases of entanglement exceeding which lie within  the conventional quantum mechanics, and also do not signal any problem.

The third  case of entanglement exceeding is the one studied in Refs. \cite{11, 12}.
 In these works, again, the system $S$ includes two qubits $A$ and $B$, and only the qubit $A$ interacts with its environment $E_A$. Choosing the initial state of the whole system-environment appropriately results that the reduced dynamics of the system $S=AB$ is given by a quantum channel, i.e., a completely positive trace-preserving map $\mathcal{E}_{AB}$. But, $\mathcal{E}_{AB}$ cannot be decomposed as $\mathcal{E}_{A}\otimes\mathcal{E}_{B}$, where  $\mathcal{E}_{A}$ and $\mathcal{E}_{B}$ are quantum channels on qubits $A$ and $B$, respectively. Therefore, the entanglement between $A$ and $B$ exceeds its initial value \cite{11, 12}.

The fourth case  has been introduced in Ref. \cite{13}, in which the system $S$ includes two  qubits $A$ and $B$, and each qubit interacts with its local environment $E_A$ and $E_B$, respectively. The reduced dynamics of the system, from some initial instant $t_i$, is given by  a map as $\Psi_{A}\otimes \Psi_{B}$, where  $\Psi_{A}$ and $\Psi_{B}$ are local non-positive maps on qubits $A$ and   $B$, respectively. Therefore, though the reduced dynamics remains local, its non-positivity results in entanglement exceeding than its initial value at $t_i$.

In this paper we will focus on this  fourth case and study  circumstances which result in  entanglement exceeding from local non-positive maps. Two such procedures will be discussed in this paper. The first procedure, as mentioned above, is the one introduced during the case studied in Ref. \cite{13}.
 In the next section, we will give general circumstances under which  this first procedure works. In addition, another example for this  procedure will be studied in Sec.~\ref{sec:C}.

The second procedure of constructing local non-positive maps which lead to entanglement exceeding and a   way for implementing them physically are given in Sec.~\ref{sec:E}. Before, in Sec.~\ref{sec:D}, we review the preliminaries needed. 

 Inaccessible entanglement, introduced in Ref. \cite{11}, refers to the supply of the  entanglement, though existed in the whole system-environment, but is inaccessible in the system alone. 
In Sec.~\ref{sec:D}, we discuss that the first procedure of entanglement exceeding studied in Sec.~\ref{sec:B}, in fact,  does not include entanglement generation. But, it only includes regaining  again in the system some amount of this supply of   entanglement.

 In addition, our proposed method for implementing physically the second  procedure of entanglement exceeding in Sec.~\ref{sec:E} is such that it includes no entanglement generation by local non-positive maps. Here again, the role of the local non-positive reduced dynamics is to make accessible in the system some amount of the  pre-existing inaccessible entanglement in the whole system-environment.

 We end our paper in Sec.~\ref{sec:F}  with a summary of our results.

\section{First procedure}\label{sec:B}

\begin{figure*}
\begin{center}
\includegraphics[width=18 cm]{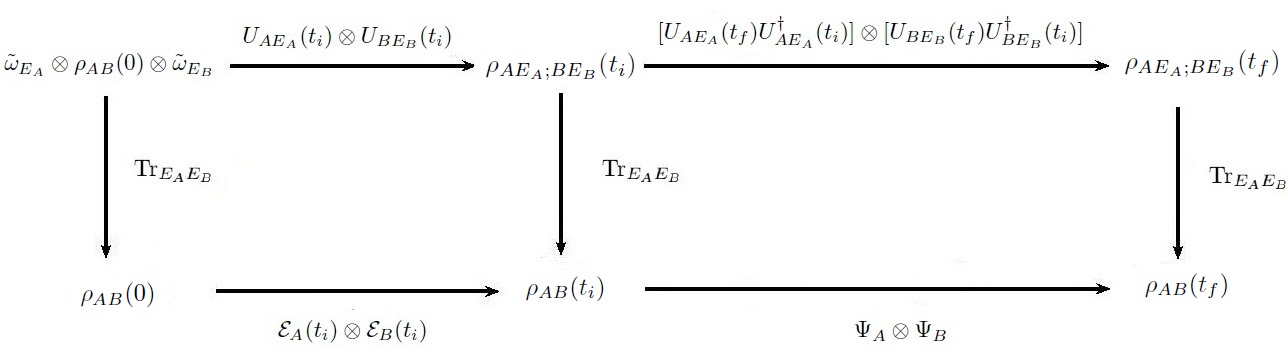}
\end{center}
\caption{At time $t=0$, the system-environment state is as $\rho_{AE_A;BE_B}(0)=\tilde{\omega}_{E_A}\otimes\rho_{AB}(0)\otimes\tilde{\omega}_{E_B}$. It evolves  unitarily as  Eq. (\ref{eq:8a}).
  So, the  initial state of the system-environment at  $t_i$ is  $ \rho_{AE_A;BE_B}(t_i)$, and its final state  at  $t_f$ is  $  \rho_{AE_A;BE_B}(t_f)$.
  Tracing over the environments $E_A$ and $E_B$ gives the reduced state of the system $S=AB$. Therefore, the state of the system at $t=0$ is 
  $   \rho_{AB}(0)   $, which is arbitrary. In addition,  the initial state of the system at $t_i$ and its final state at $t_f$ are as  
   $  \rho_{AB}(t_i)$ and   $  \rho_{AB}(t_f)$, respectively.
   According to Eq. (\ref{eq:8}),
   the  Hermitian map  $\Psi_A \otimes\Psi_B$  gives the reduced dynamics of the system  from $t_i$ to $t_f$.
    When the map $\Psi_A \otimes\Psi_B$ is non-positive,   the initial state  $  \rho_{AB}(t_i)$ is not arbitrary.}
\label{Fig3}
\end{figure*}

Consider a quantum system $S$  interacting with its environment $E$.  The whole system-environment is a closed quantum system which evolves  unitarily  \cite{14}:
\begin{equation}
\label{eq:1}
 \rho_{SE}^{\prime}= \mathrm{Ad}_U (\rho_{SE}) \equiv U \rho_{SE} U^{\dagger},
 \end{equation} 
 where $U$ is a unitary operator on $\cH_S\otimes\cH_E$.
   In addition, $\rho_{SE}$ and $\rho_{SE}^{\prime}$ are the  initial and the  final states  of the system-environment, respectively.
So, the reduced dynamics of the system is given by  
\begin{equation}
\label{eq:2}
\rho_{S}^{\prime}=\mathrm{Tr}_{E}(\rho_{SE}^{\prime})=\mathrm{Tr}_{E} \circ \mathrm{Ad}_U (\rho_{SE}).
\end{equation} 

In general, the final state of the system $\rho_{S}^{\prime}$ cannot be written as a function of its initial state $\rho_{S}=\mathrm{Tr}_{E}(\rho_{SE})$  \cite{15, 16}. But, for some restricted sets of possible initial states of the system-environment $\mathcal{S}=\lbrace\rho_{SE}\rbrace$, we have  $\rho_{S}^{\prime}=\Phi_{S}(\rho_{S})$, where $\Phi_{S}$ is a (linear  trace-preserving) Hermitian map, i.e., it maps each Hermitian operator to a Hermitian one \cite{17, 15, 18}. Restricting more the set $\mathcal{S}$, the reduced dynamics can be written as a quantum channel, i.e., $\rho_{S}^{\prime}=\mathcal{E}_{S}(\rho_{S})$, where $\mathcal{E}_S$ is a completely positive trace-preserving (CP) map \cite{19, 20, 21, 22}.

The most famous and widely used case which leads to a CP reduced dynamics is when the set of possible initial states of the system-environment is as  $\mathcal{S}=\lbrace \rho_{S}\otimes \tilde{\omega}_E \rbrace$, where $\rho_{S}$ are  arbitrary states of the system, but $ \tilde{\omega}_E $ is a fixed state of the environment \cite{14}. It is also worth noting that for each CP map $\mathcal{E}_{S}$  there exists operator sum representation as
\begin{equation}
\label{eq:3}
\begin{aligned}
\rho_{S}^{\prime}=\sum_{i}K_{i}\,\rho_{S}\,K_{i}^{\dagger},\ \ \ \sum_{i}K_{i}^{\dagger}K_{i}=I_{S},
\end{aligned}
\end{equation}
where the linear operators $K_{i}$, acting on $\cH_S$, are called the Kraus operators \cite{14}.

In this paper,  we consider a  bipartite quantum system $S=AB$ such that the parts $A$ and $B$ do not interact with each other, but each part interacts with its local  environment $E_A$ and $E_B$, respectively. So, the time evolution operator of the whole system-environment is as
\begin{equation}
\label{eq:4}
U=U_{AE_A}\otimes U_{BE_B},
\end{equation}
  where $U_{AE_A}$ and $U_{BE_B}$ are unitary operators on $\cH_A\otimes\cH_{E_A}$ and $\cH_B\otimes\cH_{E_B}$, respectively.

 We restrict our discussion to the case that the reduced dynamics of the system is given by a quantum channel  $\mathcal{E}_{S}=\mathcal{E}_{A}\otimes\mathcal{E}_{B}$, where $\mathcal{E}_{A}$ and $\mathcal{E}_{B}$ are localized CP maps acting on parts $A$ and $B$, respectively. 
This can be due to the fact that the set of possible initial states of the system-environment is factorized as
\begin{equation}
\label{eq:5}
 \mathcal{S}=\lbrace\rho_{SE}=\tilde{\omega}_{E_A}\otimes\rho_{AB}\otimes\tilde{\omega}_{E_B}\rbrace,
\end{equation}
 where $\rho_S=\rho_{AB}$ are arbitrary states on $\cH_S=\cH_A\otimes\cH_{B}$, but $\tilde{\omega}_{E_A}$ and $\tilde{\omega}_{E_B}$ are fixed states on $\cH_{E_A}$ and $\cH_{E_B}$, respectively.
  Generalizing the following discussion to the case that the reduced dynamics is given by a Hermitian map $\Phi_{S}=\Phi_{A}\otimes\Phi_{B}$ is simple.

Let's write the  dependence on the time $t$ explicitly:
\begin{equation}
\label{eq:6}
\rho_{AB}(t)=\mathcal{E}_{A}(t)\otimes\mathcal{E}_{B}(t)[\rho_{AB}(0)].
\end{equation} 
Assume that, for some time $t_i> 0$, there exist (linear trace-preserving) maps $\mathcal{E}_{A}^{-1}(t_i)$  and $\mathcal{E}_{B}^{-1}(t_i)$ such that $\mathcal{E}_{A}^{-1}(t_i)\circ \mathcal{E}_{A}(t_i)= \mathrm{id}_A$ and  $\mathcal{E}_{B}^{-1}(t_i)\circ \mathcal{E}_{B}(t_i)= \mathrm{id}_B$, where $\mathrm{id}_A$ and $\mathrm{id}_B$ are identity maps on  $\mathcal{L}(\cH_{A})$ and $\mathcal{L}(\cH_{B})$, respectively. ($\mathcal{L}(\cH)$ is the vector space of linear operators on $\cH$.)
Note that the inverse map $\mathcal{E}_{A}^{-1}(t_i)$ ($\mathcal{E}_{B}^{-1}(t_i)$) exists, if and only if the quantum channel $\mathcal{E}_{A}(t_i)$ ($\mathcal{E}_{B}(t_i)$) is one to one, i.e.,  it maps different initial states to different final ones. To check whether a linear map $\mathcal{E}$ is one to one or not, it is enough to check whether it maps a set of linearly independent operators, which constructs a basis for $\mathcal{L}(\cH)$, to (another) set of linearly independent operators or not.

When the inverse maps at time $t_i$ exist, we have
\begin{equation}
\label{eq:7}
\rho_{AB}(0)=\mathcal{E}_{A}^{-1}(t_i)\otimes\mathcal{E}_{B}^{-1}(t_i)[\rho_{AB}(t_i)].
\end{equation}
So, using Eqs. (\ref{eq:6}) and (\ref{eq:7}), we conclude that (for  $t_f \geq t_i$)
\begin{equation}
\label{eq:8}
\begin{aligned}
 \rho_{AB}(t_f)=\mathcal{E}_{A}(t_f)\otimes\mathcal{E}_{B}(t_f)[\rho_{AB}(0)] \qquad\qquad \qquad\qquad \qquad \ \\
  =\left( \mathcal{E}_{A}(t_f)\circ \mathcal{E}_{A}^{-1}(t_i)\right)  \otimes \left( \mathcal{E}_{B}(t_f)\circ \mathcal{E}_{B}^{-1}(t_i)\right) [\rho_{AB}(t_i)]   \\
=\Psi_A \otimes \Psi_B [\rho_{AB}(t_i)],\qquad\qquad \qquad\qquad\qquad\qquad \ 
\end{aligned}
\end{equation} 
where $\Psi_A=\mathcal{E}_{A}(t_f)\circ \mathcal{E}_{A}^{-1}(t_i)$ and $\Psi_B=\mathcal{E}_{B}(t_f)\circ \mathcal{E}_{B}^{-1}(t_i)$ are (linear  trace-preserving) Hermitian maps on  $\mathcal{L}(\cH_{A})$ and $\mathcal{L}(\cH_{B})$, respectively.

In Eq. (\ref{eq:8}), the reduced dynamics of the bipartite system $S=AB$ from $t_i$ to $t_f$ is given by a localized map.
But, since  $\Psi_A$ and $\Psi_B$ are not CP in general, we may encounter unexpected entanglement dynamics, i.e., 
$\rho_{AB}(t_f)$ may be more entangled than $\rho_{AB}(t_i)$, as has been seen in the case studied in Ref. \cite{13}. In the next section, we will give another such example. 

When $\Psi_A \otimes\Psi_B$ is not a positive map, it clearly cannot give correctly the reduced dynamics of the system $S=AB$ for arbitrary initial state of the system. In Fig. \ref{Fig3}, we illustrated that for which set of initial states we use the Hermitian map  $\Psi_A \otimes\Psi_B$ to describe the reduced dynamics.

First, remember that at time $t=0$  the set of  possible system-environment states is as Eq. (\ref{eq:5}), i.e.,  $\mathcal{S}(0)=\lbrace \tilde{\omega}_{E_A}\otimes\rho_{AB}(0)\otimes\tilde{\omega}_{E_B}\rbrace$, where $\rho_{AB}(0)$ are arbitrary. The whole system-environment evolves under the unitary operator  $U_{AE_A}(t )\otimes U_{BE_B}(t )$, which is as  Eq. (\ref{eq:4}). Therefore, the system-environment state at time $t$ is given by
\begin{equation}
\label{eq:8a}
\begin{aligned}
 \rho_{AE_A;BE_B}(t)=   \qquad\qquad\qquad\qquad\qquad\qquad\qquad\qquad\qquad      \\
  (\mathrm{Ad}_{U_{AE_A}(t)} \otimes \mathrm{Ad}_{U_{BE_B}(t)}) [\tilde{\omega}_{E_A}\otimes\rho_{AB}(0)\otimes\tilde{\omega}_{E_B}].  
\end{aligned}
\end{equation} 
 So, the set of  initial states of the system-environment at  $t_i$ is given by $\mathcal{S}(t_i)=\lbrace  \rho_{AE_A;BE_B}(t_i)\rbrace$, and the set of final states of the system-environment at  $t_f$ is as  $\mathcal{S}(t_f)=\lbrace  \rho_{AE_A;BE_B}(t_f)\rbrace$.
 
 Next, note that the set of the system states at time $t$ is given by $\mathcal{S}_S(t)= \mathrm{Tr}_{E_AE_B} \mathcal{S}(t)$. Therefore, the set of  possible states of the system at $t=0$ is $\mathcal{S}_S(0)=\lbrace   \rho_{AB}(0)  \rbrace = \mathcal{D}_S$, where $\mathcal{D}_S$ is the set of all   density operators on $\cH_S=\cH_A\otimes\cH_B$. In other words, the state of the system $S=AB$ at $t=0$ is arbitrary.
  In addition, the set of initial states of the system  at  $t_i$ is given by  $\mathcal{S}_S(t_i)=\lbrace  \rho_{AB}(t_i)= \mathrm{Tr}_{E_AE_B} (\rho_{AE_A;BE_B}(t_i)) \rbrace$, and the set of final states of the system  at  $t_f$ is as $\mathcal{S}_S(t_f)=\lbrace  \rho_{AB}(t_f)=\mathrm{Tr}_{E_AE_B} (\rho_{AE_A;BE_B}(t_f)) \rbrace$. 
 
 The  Hermitian map  $\Psi_A \otimes\Psi_B$ in Eq. (\ref{eq:8}) gives the reduced dynamics of the system $S=AB$, from $t_i$ to $t_f$, for initial states of the system $\rho_{AB}(t_i) \in \mathcal{S}_S(t_i)$. When the map $\Psi_A \otimes\Psi_B$ is non-positive, then, surely, the set $\mathcal{S}_S(t_i)$ is a proper subset of  $\mathcal{D}_S$.

Finally, note that the dynamics of the whole system-environment from $t_i$ to $t_f$ is given by the unitary operator $[U_{AE_A}(t_f)U_{AE_A}^\dagger(t_i)]\otimes [U_{BE_B}(t_f)U_{BE_B}^\dagger(t_i)]$  which is factorized as Eq. (\ref{eq:4}). But, in general, the initial states $ \rho_{AE_A;BE_B}(t_i)$ are not factorized as $ \rho_{AE_A;BE_B}(0)= \tilde{\omega}_{E_A}\otimes\rho_{AB}(0)\otimes\tilde{\omega}_{E_B}$. Therefore,  $\Psi_A$ and $\Psi_B$ are not CP in general. In other words, the non-CP-ness of  $\Psi_A$ or $\Psi_B$ implies  that the set of initial states of the system-environment  $\mathcal{S}(t_i)$ includes correlated states.

\section{Example for the first procedure}\label{sec:C}  
 
In this section, we consider the case studied in Ref. \cite{5} to illustrate the results of the previous section. They considered a system $S$ including two qubits $A$ and $B$. Each qubit is a two-level atom interacting with its local environment (reservoir) constructed from the quantized modes of the electromagnetic field in a cavity. 
So, the dynamics of the whole system-environment is as Eq. (\ref{eq:4}). In addition, since the initial state of the 
 system-environment at $t=0$ is factorized as Eq. (\ref{eq:5}), the reduced dynamics of the two-qubit system is localized as Eq. (\ref{eq:6}).

If we consider an initial state of the qubit $A$ as 
\begin{equation}
\label{eq:9}
\rho_A(0)=\left( 
\begin{matrix} 
\rho_A^{00}(0) & \ \rho_A^{01}(0) \\ \\
\rho_A^{10}(0) & \ \rho_A^{11}(0)
\end{matrix}
\right), 
\end{equation} 
 then the quantum channel $\mathcal{E}_{A}(t)$ maps it to the final state \cite{5}
\begin{equation}
\label{eq:10}
\begin{aligned}
\rho_A(t)= \mathcal{E}_{A}(t)[\rho_A(0)] \qquad\qquad\qquad\qquad\qquad\qquad\qquad\qquad \\
= \left( 
\begin{matrix} 
 \rho_A^{00}(0)+ \rho_A^{11}(0)(1-P(t)) &  \qquad \rho_A^{01}(0)\sqrt{P(t)} \\ \\
\rho_A^{10}(0)\sqrt{P(t)} & \ \rho_A^{11}(0)P(t) 
\end{matrix}
\right). 
\end{aligned}
\end{equation}  
 In the strong coupling regime, the function $P(t)$ is as
\begin{equation}
\label{eq:11}
P(t)=e^{- \lambda t} \left[  \cos (\frac{tD}{2}) + \frac{\lambda}{D} \sin (\frac{tD}{2}) \right]^2, 
\end{equation} 
where $D=\sqrt{2 \gamma_0 \lambda -\lambda^2}$, and  the positive constants $\gamma_0$ and $\lambda$ are  related to the coupling between the atom and the reservoir and to the width of the spectral density of the reservoir, respectively  \cite{5}. The function $P(t)$ has discrete zeros at
\begin{equation}
\label{eq:12}
t_n=\frac{2}{D} \left[ n \pi -\arctan (\frac{D}{\lambda})\right], 
\end{equation}
where $n=1, 2, 3, \dots $.  The quantum channel $\mathcal{E}_{B}(t)$ acts similarly as Eq. (\ref{eq:10}) on each initial state $\rho_B(0)$ of qubit $B$.
 
Note that each arbitrary operator $O \in \mathcal{L}(\cH_A)$ can be written as a linear combination of states $\rho_A$. In addition,
  the quantum channel $\mathcal{E}_{A}(t)$ is a linear map. Therefore, its action on arbitrary operator $O \in \mathcal{L}(\cH_A)$ is also as Eq. (\ref{eq:10}). So, the linearly independent set 
\begin{equation}
\label{eq:13}
\begin{aligned}
X=\left( 
\begin{matrix} 
0 & 1 \\
1 & 0
\end{matrix}
\right), \quad  
Y=\left( 
\begin{matrix} 
0 & -i \\
i & 0
\end{matrix}
\right), \\  
Z=\left( 
\begin{matrix} 
1 & 0 \\
0 & -1
\end{matrix}
\right), \quad 
I=\left( 
\begin{matrix} 
1 & 0 \\
0 & 1
\end{matrix}
\right), 
\end{aligned}
\end{equation}  
are mapped by $\mathcal{E}_{A}(t)$ to the set
 \begin{equation}
\label{eq:14}
\begin{aligned}
X(t)=\mathcal{E}_{A}(t)[X]=\left( 
\begin{matrix} 
0 & \sqrt{P(t)}  \\
\sqrt{P(t)}  & 0
\end{matrix}
\right)=\sqrt{P(t)}X, \qquad \\  
Y(t)=\mathcal{E}_{A}(t)[Y]=\left( 
\begin{matrix} 
0 & -i \sqrt{P(t)} \\
i \sqrt{P(t)} & 0
\end{matrix}
\right)= \sqrt{P(t)}Y, \quad  \\  
Z(t)=\mathcal{E}_{A}(t)[Z]=\left( 
\begin{matrix} 
 P(t) & 0 \\
0 & -P(t)
\end{matrix}
\right)= P(t)Z, \qquad\qquad\quad \\ 
I(t)=\mathcal{E}_{A}(t)[I]=\left( 
\begin{matrix} 
2-P(t) & 0 \\
0 &  P(t)
\end{matrix}
\right)=I+(1-P(t))Z.
\end{aligned}
\end{equation}
 The above set is linearly independent if and only if $P(t) \neq 0$. Therefore, for all $t \neq t_n$ in Eq. \eqref{eq:12}, the map $\mathcal{E}_{A}(t)$ is invertible, and so is $\mathcal{E}_{B}(t)$. 
 Consequently, for $t_i \neq t_n$ in Eq. \eqref{eq:12}, we can write Eqs.  \eqref{eq:7} and  \eqref{eq:8}.
 
In addition, when $P(t) \neq 0$,   Eq.   \eqref{eq:14} results that
\begin{equation}
\label{eq:14a}
\begin{aligned}
\mathcal{E}_{A}^{-1}(t)[X]=\frac{1}{\sqrt{P(t)}}X, \qquad \\
\mathcal{E}_{A}^{-1}(t)[I]=I+ \frac{P(t)-1}{P(t)}Z.
\end{aligned}
\end{equation}
So, a state of the qubit $A$ as 
\begin{equation}
\label{eq:14b}
\begin{aligned}
\rho_A=\frac{1}{2}(I+r_xX),
\end{aligned}
\end{equation}
where $-1 \leq r_x \leq 1$, is mapped by $\Psi_A=\mathcal{E}_{A}(t_f)\circ \mathcal{E}_{A}^{-1}(t_i)$ to
\begin{equation}
\label{eq:14c}
\begin{aligned}
\rho^\prime_A=\Psi_A(\rho_A)=\frac{1}{2}(I+r_x\sqrt{\frac{P(t_f)}{P(t_i)}} X+r^\prime_zZ),
\end{aligned}
\end{equation}
where $r^\prime_z$ is a real coefficient (which its value is not important for us).
When $\frac{P(t_f)}{P(t_i)} > 1$, for  $\sqrt{\frac{P(t_i)}{P(t_f)}} < \vert r_x  \vert \leq 1$, we have $\lvert r_x\sqrt{\frac{P(t_f)}{P(t_i)}}\rvert > 1 $. That is, for this range of $r_x$, the operator $\rho^\prime_A$ in Eq.  \eqref{eq:14c} is not a valid density operator. Therefore, when $\frac{P(t_f)}{P(t_i)} > 1$, the Hermitian map $\Psi_A$ is a non-positive map. A similar discussion  can be given for the map
 $\Psi_B=\mathcal{E}_{B}(t_f)\circ \mathcal{E}_{B}^{-1}(t_i)$ too.

\begin{figure}
\begin{center}
\includegraphics[width=8.7 cm]{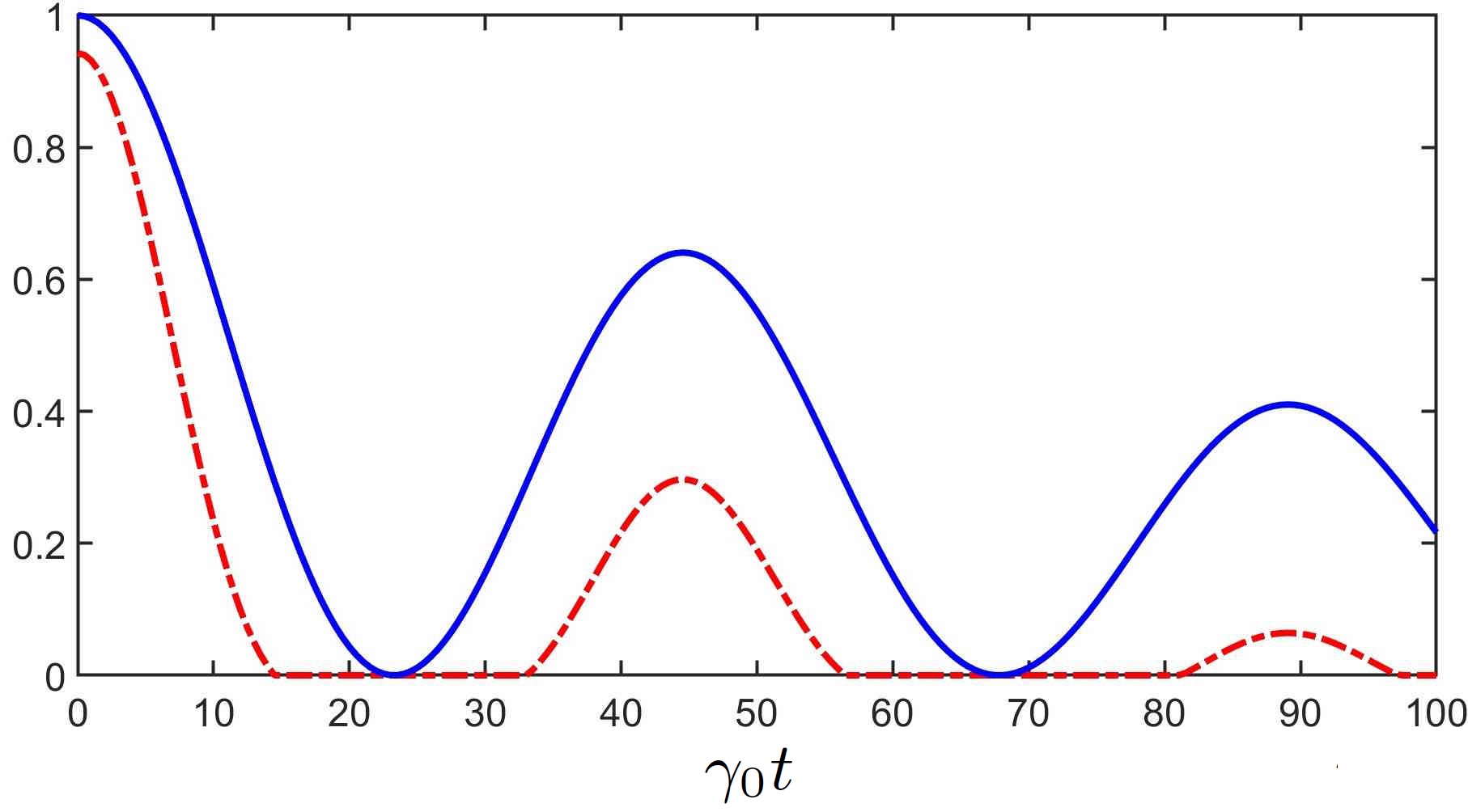}
\end{center}
\caption{Concurrence (red
dash-dotted line) for $\alpha=\frac{1}{\sqrt{3}}$, and  $P(t)$ in Eq. \eqref{eq:11} (blue solid line) as
the functions of the scaled time $\gamma_0 t$. In addition, $\lambda=0.01 \gamma_0$.}
\label{Fig1}
\end{figure}

 In Ref. \cite{5} the concurrence \cite{22a, 27} is used as the measure of the entanglement. When the initial state of the two-qubit system $S=AB$ is as $\rho_{AB}(0)=\vert \psi_{AB} \rangle \langle \psi_{AB} \vert$, where 
  $\vert \psi_{AB} \rangle=\alpha \vert 0_A \rangle\vert 0_B\rangle + \beta \vert 1_A \rangle\vert 1_B\rangle$,  $\alpha$ is real and $\alpha^2+ \vert\beta\vert^2=1$, the concurrence of the final state $\rho_{AB}(t)$ in Eq. \eqref{eq:6} is given by \cite{5}
\begin{equation}
\label{eq:14d}
C(t)= \max \left\lbrace 0, 2 \sqrt{1-\alpha^2}P(t)[\alpha- \sqrt{1-\alpha^2} \left( 1-P(t)\right) ]\right\rbrace.
\end{equation} 
In Fig. \ref{Fig1}, we plotted  $C(t)$  for $\alpha=\frac{1}{\sqrt{3}}$, as  Fig. 4 of Ref. \cite{5}, and also
 $P(t)$ in Eq. \eqref{eq:11}.   
 We choose $t_i \gamma_0 =20$ which is an appropriate initial instant, since $P(t_i) \neq 0$  and so $\mathcal{E}^{-1}_{A}(t_i)$ and  $\mathcal{E}^{-1}_{B}(t_i)$ exist. From Fig. \ref{Fig1}, we see that for this $t_i$ the state $\rho_{AB}(t_i)$ is  separable. Then, if we choose, e.g.,  $t_f \gamma_0 =40$, we see that $\rho_{AB}(t_f)$ is an entangled state. So, the local Hermitian map $\Psi_A \otimes \Psi_B$ in Eq.  \eqref{eq:8} is entanglement generating.

 In the case studied in this section, $\mathcal{E}_{A}$  and  $\mathcal{E}_{B}$ are the same. So, we can call them both as $\mathcal{E}$. Similarly, we can call the both 
  $\Psi_A$ and  $\Psi_B$ as $\Psi$.
Note that $\Psi$  is a non-positive map, since, for $t_i=\frac{20}{\gamma_0}$ and $t_f=\frac{40}{\gamma_0}$, we have $\frac{P(t_f)}{P(t_i)} > 1$.

In summary, the non-positive local map $\Psi \otimes \Psi$ can result in entanglement generating
 (exceeding).

 From Eqs. \eqref{eq:6} and \eqref{eq:8}, we have
\begin{equation}
\label{eq:a14}
\begin{aligned}
 \rho_{AB}(t_f)=\mathcal{E} (t_f)\otimes\mathcal{E} (t_f)[\rho_{AB}(0)] \qquad\qquad \ \   \\ 
=\Psi  \otimes \Psi  [\rho_{AB}(t_i)] \qquad\qquad\qquad\quad   \ \ \\
=( \Psi   \circ  \mathcal{E} (t_i)) \otimes( \Psi   \circ  \mathcal{E} (t_i))[\rho_{AB}(0)].
\end{aligned}
\end{equation}
Since $ \rho_{AB}(0)$ is an arbitrary state in $\mathcal{D}_S$, we conclude that $\mathcal{E} (t_f)$ is divisible as  $\mathcal{E} (t_f)=\Psi   \circ  \mathcal{E} (t_i)$. If $\Psi$ is CP (positive), then $\mathcal{E} (t_f)$ is CP-divisible (P-divisible), at least when we choose the intermediate time between $0$ and $t_f$ as $t_i$.
 CP-divisibility has been proposed as a possible definition of the Markovianity of a quantum operation \cite{25a, 26a}.

When $\Psi$ is CP, then $\rho_{AB}(t_f)$ cannot be more entangled than $\rho_{AB}(t_i)$. In addition, when $\rho_{AB}(t_i)$ is separable and  $\Psi$ is positive, performing $\Psi  \otimes \Psi $ on $\rho_{AB}(t_i)$ cannot map it to the entangled state $\rho_{AB}(t_f)$. Therefore,   the entanglement dynamics in Fig. \ref{Fig1},  from $t_i=\frac{20}{\gamma_0}$ to $t_f=\frac{40}{\gamma_0}$, results that  $\Psi$ is non-positive, and so the quantum channel $\mathcal{E} (t_f)$ is non-Markovian. More precisely,  $\mathcal{E} (t_f)$ is non-P-divisible.

We end this section with the following additional point.   Fig. \ref{Fig1}   shows that  local non-positive maps sometimes act as we expect, i.e., they sometimes  do not generate entanglement.  For example, we again choose  $t_i \gamma_0 =20$, but we choose  $t_f \gamma_0 =75$. So,  $\frac{P(t_f)}{P(t_i)} > 1$, and  therefore  the map $\tilde{\Psi}=\mathcal{E}(t_f)\circ \mathcal{E}^{-1}(t_i)$ is non-positive too. We see that the both states $\rho_{AB}(t_i)$  and $\rho_{AB}(t_f)$ are  separable, i.e., the non-positive local map $\tilde{\Psi} \otimes \tilde{\Psi}$ cannot map the separable state $\rho_{AB}(t_i)$ to an entangled one.

 \section{Entanglement breaking and entanglement annihilating channels; inaccessible entanglement}\label{sec:D} 
 
 To introduce the second procedure of entanglement exceeding in the next section, we need some results on 
 entanglement breaking and entanglement annihilating channels. So, we review these concepts in the following.
 
 A quantum channel $\mathcal{E}_{A}$ is called entanglement breaking if $\mathcal{E}_{A}\otimes \mathrm{id}_B [\rho_{AB}]$, for arbitrary bipartite state $\rho_{AB}$, is always separable \cite{23, 24}.
In other words,  $\mathcal{E}_{A}\otimes \mathrm{id}_B$ maps all entangled states to separable ones. 
 
 An important theorem proven in Ref. \cite{23} is that a quantum channel $\mathcal{E}_{A}$ is entanglement breaking if and only if 
 $\mathcal{E}_{A}\otimes \mathrm{id}_B [\phi^+_{AB}]$ is a separable state, where $\phi^+_{AB}=\vert\phi_{AB}^{+}\rangle\langle\phi_{AB}^{+}\vert$, and $\vert\phi_{AB}^{+}\rangle=\frac{1}{\sqrt{d}} \sum_{j=1}^d \vert j_{A}\rangle \vert j_{B}\rangle$ is the maximally entangled state. (The both subsystems $A$ and $B$ are $d$-dimensional.)
 
A  quantum channel $\mathcal{E}$ is called entanglement annihilating if $\mathcal{E}\otimes \mathcal{E} [\rho_{AB}]$, for arbitrary bipartite state $\rho_{AB}$, is always separable \cite{25, 26}. 
When $\mathcal{E}$ is entanglement breaking, then it is  entanglement annihilating too, since $\mathcal{E}\otimes \mathcal{E} [\rho_{AB}]=(\mathcal{E}\otimes \mathrm{id}) \circ (\mathrm{id}\otimes\mathcal{E})[\rho_{AB}]$, which is  obviously separable. But, the  inverse implication is not true, in general, i.e., a quantum channel $\mathcal{E}$  can be entanglement annihilating while it is not entanglement breaking \cite{25, 26}. We will use this fact in the next section.

Another concept which we will use is the concept of inaccessible entanglement, introduced in Ref. \cite{11}.
There, as the current work, we considered the case that 
the two parts $A$ and $B$ of our bipartite system $S=AB$ are separated from each other and each part interacts with its  local environment $E_{A}$ and  $E_{B}$, respectively. 
So, our quadripartite configuration consists of two separated parts $AE_{A}$ and $BE_{B}$.
Inaccessible entanglement is defined as \cite{11}
\begin{equation}
\label{eq:15}
\mathcal{M}_{I}=\mathcal{M}(\rho_{AE_{A};BE_{B}})-\mathcal{M}(\rho_{AB}),
\end{equation} 
where $\mathcal{M}$ is an appropriate entanglement measure (monotone) for bipartite systems, $\rho_{AE_{A};BE_{B}}$ is the whole state of the system-environment 
 and $\rho_{AB}=\mathrm{Tr_{E_{A}E_{B}}}(\rho_{AE_{A};BE_{B}})$ is the reduced state of the system.
 In addition, $\mathcal{M}(\rho_{AE_{A};BE_{B}})$ is calculated according to the  bipartition $({AE_{A}; BE_{B}})$.
 
Each entanglement measure $\mathcal{M}$ is zero for all separable states (and some entangled states) and is positive for (other) entangled ones. An important property of an entanglement measure (monotone) is that it does not increase under local CP operations \cite{27}. Partial traces over $E_{A}$ and $E_{B}$ are  local CP operations \cite{14,28}.
 Therefore, we always have
  $\mathcal{M}(\rho_{AE_{A};BE_{B}})\geq\mathcal{M}(\rho_{AB})$, and so  $\mathcal{M}_{I}\geq 0$ in Eq. \eqref{eq:15}.
  
  Assuming that we have access only to the system $S=AB$ and not to the environments, the meaning of the inaccessible entanglement is clear: It measures the amount of entanglement which is present between the two separated parts $AE_{A}$ and $BE_{B}$, but is inaccessible for us.
  
Now, let's explain more some previously stated results. 
First, note that a time evolution operator $U$ as Eq. \eqref{eq:4} is  a local CP operation according to the  bipartition $({AE_{A}; BE_{B}})$.  Its   inverse, i.e., the operator $U^\dagger$, is  a local CP operation too. Therefore, implementing such $U$ in Eq. \eqref{eq:1}, we have $\mathcal{M}(\rho_{AE_{A};BE_{B}})\geq\mathcal{M}(\rho^\prime_{AE_{A};BE_{B}})$ and $\mathcal{M}(\rho^\prime_{AE_{A};BE_{B}}) \geq \mathcal{M}(\rho_{AE_{A};BE_{B}})$, simultaneously.  We conclude that, for each $U$  as Eq. \eqref{eq:4}, 
\begin{equation}
\label{eq:16}
\begin{aligned}
 \mathcal{M}(\rho_{AE_{A};BE_{B}})=\mathcal{M}(\rho^\prime_{AE_{A};BE_{B}}).
\end{aligned}
\end{equation} 
Similarly, adding local environments and its inverse, i.e., discarding (tracing over) them, are both  local CP maps.
 So, when $\rho_{AE_{A};BE_{B}}=\tilde{\omega}_{E_A}\otimes\rho_{AB}\otimes\tilde{\omega}_{E_B}$ as Eq. \eqref{eq:5}, we have
\begin{equation}
\label{eq:17}
\begin{aligned}
 \mathcal{M}(\rho_{AE_{A};BE_{B}})=\mathcal{M}(\rho_{AB}).
\end{aligned}
\end{equation} 
On the other hand, as stated before, partial traces over $E_{A}$ and $E_{B}$ are  local CP operations. Therefore, 
\begin{equation}
\label{eq:18}
\begin{aligned}
 \mathcal{M}(\rho^\prime_{AE_{A};BE_{B}})\geq\mathcal{M}(\rho^\prime_{AB}),
\end{aligned}
\end{equation} 
where  $\rho^\prime_{AB}=\mathrm{Tr_{E_{A}E_{B}}}(\rho^\prime_{AE_{A};BE_{B}})$ is the final state of the system.
Using Eqs. \eqref{eq:16} to \eqref{eq:18}, we conclude that $\mathcal{M}(\rho^\prime_{AB}) \leq \mathcal{M}(\rho_{AB})$. This is why the increase of the entanglement in the case studied in Ref. \cite{8} is a signal of a problem, as mentioned in the Introduction.

 Second, for the case studied in Sec.~\ref{sec:B}, again since the time evolution is as Eq.  \eqref{eq:4}, and in Eq. \eqref{eq:5} we have $\rho_{AE_{A};BE_{B}}(0)=\tilde{\omega}_{E_A}\otimes\rho_{AB}(0)\otimes\tilde{\omega}_{E_B}$, we conclude that 
\begin{equation}
\label{eq:19}
\begin{aligned}
\mathcal{M}(\rho_{AB}(0))= \mathcal{M}(\rho_{AE_{A};BE_{B}}(0))=\mathcal{M}(\rho_{AE_{A};BE_{B}}(t)),
\end{aligned}
\end{equation} 
 using Eqs. \eqref{eq:16} and \eqref{eq:17}. In other words, the amount of the entanglement existing between the two distinct parts $AE_{A}$ and $BE_{B}$ remains unchanged during the time evolution under the localized unitary operator as Eq.  \eqref{eq:4}, and it is equal to   the amount of entanglement existing in $\rho_{AB}(0)$, the  state of the system at $t=0$.
 
Now, consider the example studied in  Sec.~\ref{sec:C}.  At the initial instant $t_i=\frac{20}{\gamma_0}$, the state  $\rho_{AB}(t_i)$ is separable. This means that the amount of the inaccessible entanglement at this time $t_i$, i.e., 
\begin{equation}
\begin{aligned}
\label{eq:20}
\mathcal{M}_{I}(t_i)=\mathcal{M}(\rho_{AE_{A};BE_{B}}(t_i))-\mathcal{M}(\rho_{AB}(t_i))  \\
=\mathcal{M}(\rho_{AB}(0))-0  \qquad \qquad\qquad\quad \  \\
=\mathcal{M}(\rho_{AB}(0)),   \qquad \qquad\qquad\qquad \quad
\end{aligned}
\end{equation} 
is the maximum.
During the time evolution from $t_i$ to $t_f=\frac{40}{\gamma_0}$, some amount of this inaccessible entanglement becomes accessible for us and  $\rho_{AB}(t_f)$ becomes entangled. Emphasizing again, the amount of the entanglement  between the two distinct parts $AE_{A}$ and $BE_{B}$ has always the fixed value $\mathcal{M}(\rho_{AB}(0))$. During the time evolution, some amount of it  becomes accessible and the other remains inaccessible.

In summary, strictly speaking, the localized  non-positive map $\Psi_A \otimes \Psi_B$ in Eq. \eqref{eq:8} does not generate entanglement between $A$ and $B$. It only makes some amount of the pre-existing, but inaccessible, entanglement between  $AE_{A}$ and $BE_{B}$ accessible for us, i.e.,  accessible in the system $S=AB$.

 \section{ Second procedure }\label{sec:E}

 Consider a quantum channel $\mathcal{E}$ which has the three following  properties, simultaneously:
 
 (1) It is invertible.
 
 (2) It is entanglement annihilating.
 
 (3) But, it is not entanglement breaking.
 \\
So, implementing $\mathrm{id}\otimes \mathcal{E}$ on some entangled states $\rho^{(en)}_{AB}$, including the maximally entangled state $\phi^+_{AB}$, results in entangled states, while  $ \mathcal{E}\otimes \mathcal{E}[\rho^{(en)}_{AB}]$ are separable. Now, we have
\begin{equation}
\label{eq:21}
\begin{aligned}
\mathcal{E}^{-1}\otimes \mathrm{id} \left[ \mathcal{E}\otimes\mathcal{E} [\rho^{(en)}_{AB}] \right] 
=\mathrm{id}\otimes \mathcal{E} [\rho^{(en)}_{AB}],
\end{aligned}
\end{equation} 
i.e., performing the localized map $\mathcal{E}^{-1}\otimes \mathrm{id}$ on the separable state $ \mathcal{E}\otimes \mathcal{E}[\rho^{(en)}_{AB}]$ results in the entangled state $\mathrm{id}\otimes \mathcal{E} [\rho^{(en)}_{AB}]$. Note that $\mathcal{E}^{-1}$ is a (linear trace-preserving) non-positive map.
 Otherwise, acting $\mathcal{E}^{-1}\otimes \mathrm{id}$ on each separable state $\rho_{AB}^{(sep)}= \sum_j p_j \rho_A^{(j)} \otimes \rho_B^{(j)}$ will result in a  separable one.
Eq. \eqref{eq:21} provides us the second procedure of entanglement generating (exceeding) using the non-positive local  map $\mathcal{E}^{-1}\otimes \mathrm{id}$.

As an example, we consider the generalized amplitude damping (GAD) channel  \cite{14, 24, 29}. 
It is a quantum channel acting on a qubit with the following Kraus operators:
\begin{equation}
\label{eq:22}
\begin{aligned}
K_1=\sqrt{1-n}\left( 
\begin{matrix} 
1 & 0 \\
0 & \ \sqrt{1-\gamma}
\end{matrix}
\right), \\ 
K_2=\sqrt{1-n} \left( 
\begin{matrix} 
0 & \ \sqrt{\gamma} \\
0 & 0
\end{matrix}
\right),  \qquad \\  
K_3=\sqrt{n}\left( 
\begin{matrix} 
 \sqrt{1-\gamma} & \ 0 \\
0 & 1
\end{matrix}
\right), \qquad  \\
K_4=\sqrt{n} \left( 
\begin{matrix} 
0 & \ 0 \\
\sqrt{\gamma} & 0
\end{matrix}
\right), \qquad\qquad
\end{aligned}
\end{equation}  
where   $\gamma \in [0, 1]$ is the amplitude damping parameter and $n \in [0, 1]$ is related to the environment temperature.
Let's denote the GAD channel as $\mathcal{E}(\gamma , n)$. 
It can be shown that \cite{29}
\begin{equation}
\label{eq:23}
\begin{aligned}
\mathcal{E}(\gamma , n)[X]= \sqrt{1-\gamma} X,  \qquad   \\
\mathcal{E}(\gamma , n)[Y]=\sqrt{1-\gamma} Y, \qquad  \ \\
\mathcal{E}(\gamma , n)[Z]=(1-\gamma) Z,   \qquad  \  \\
\mathcal{E}(\gamma , n)[I]=I+\gamma (1-2n) Z .
\end{aligned}
\end{equation}
So, the linearly independent set $\left\lbrace X, Y, Z, I\right\rbrace $ is mapped to a linearly independent set
if and only if $\gamma \neq 1$. Therefore, the GAD channel
$\mathcal{E}(\gamma , n)$ is invertible if and only if $\gamma \neq 1$.

\begin{figure}
\begin{center}
\includegraphics[width=8.5 cm]{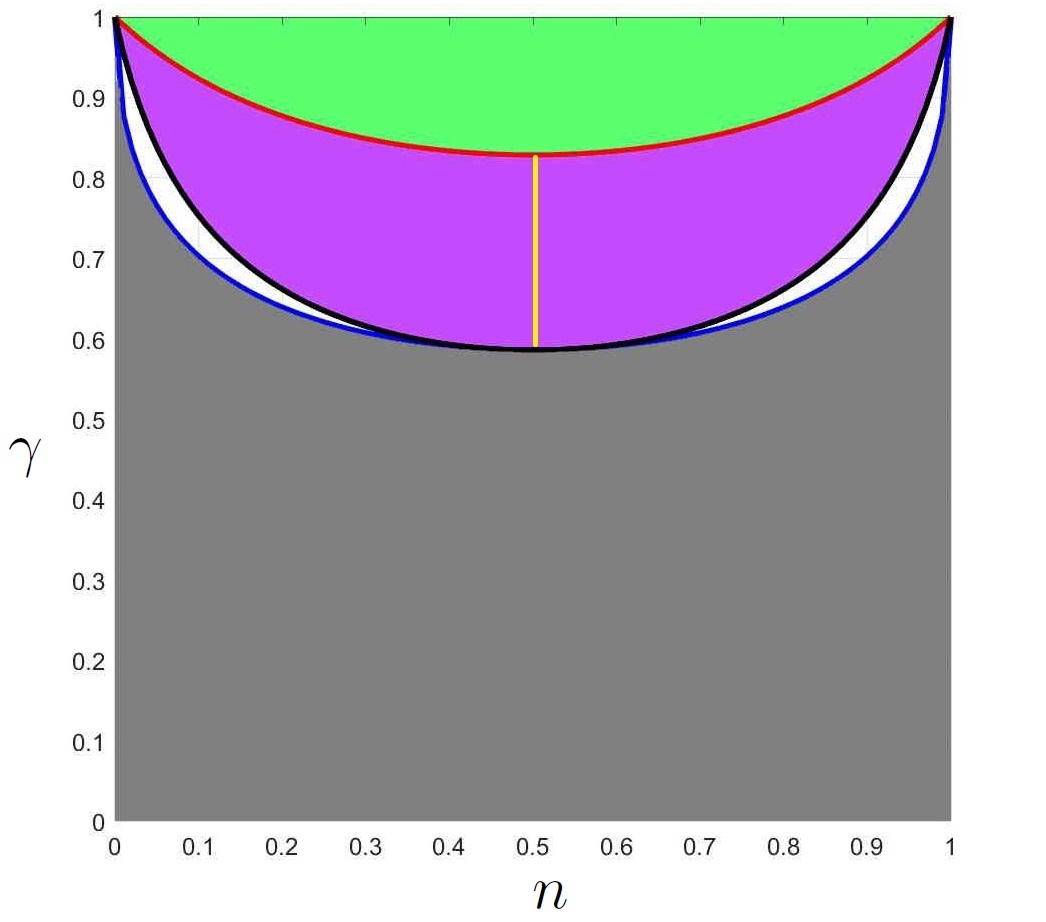}
\end{center}
\caption{Regions  of parameters $ \gamma$ and $n$ for which the GAD channel $\mathcal{E}(\gamma , n)$ is entanglement breaking or entanglement annihilating: In the green region,  $\mathcal{E}(\gamma , n)$ is entanglement breaking. On the straight yellow line, for which $n=\frac{1}{2}$, the channel  $\mathcal{E}(\gamma , n)$ is  entanglement annihilating, according to the analytical result of Ref. \cite{25}. In the violet region,  $\mathcal{E}(\gamma , n)$ is entanglement annihilating, according to the numerical result of Ref. \cite{30}.}
\label{Fig4}
\end{figure}

As stated in Sec.~\ref{sec:D},  the range of the parameters $\gamma$ and $n$ for which $\mathcal{E}(\gamma , n)$ is  entanglement breaking can be determined simply by checking whether $\mathcal{E}(\gamma , n)\otimes \mathrm{id} [\phi^+_{AB}]$ is separable or not. For $\gamma \geq \frac{\sqrt{1+4n(1-n)}-1}{2n(1-n)}$, the state 
 $\mathcal{E}(\gamma , n)\otimes \mathrm{id} [\phi^+_{AB}]$ is separable \cite{25}, and so the channel  $\mathcal{E}(\gamma , n)$ is  entanglement breaking. This region is illustrated as the green region in Fig. \ref{Fig4}. (The red line in Fig. \ref{Fig4} represents the curve $\gamma = \frac{\sqrt{1+4n(1-n)}-1}{2n(1-n)}$.)

In addition, it has been shown analytically  \cite{25} and numerically  \cite{25, 30} that the region in the $(\gamma, n)$-plane for which $\mathcal{E}(\gamma , n)$ is  entanglement annihilating does not coincides with the region for which it is  entanglement breaking. As expected, the  entanglement annihilating region includes the 
entanglement breaking one, but it is larger than the entanglement breaking region, as can be seen in  Fig.  \ref{Fig4}.

For $\gamma \leq \frac{1-\sqrt{2n(1-n)}}{1-2n(1-n)}$, the state  $\mathcal{E}(\gamma , n)\otimes \mathcal{E}(\gamma , n) [\phi^+_{AB}]$ is entangled \cite{25}. The curve  $\gamma = \frac{1-\sqrt{2n(1-n)}}{1-2n(1-n)}$ is illustrated in blue in  Fig.  \ref{Fig4}. Therefore, in the gray region in  Fig.  \ref{Fig4}, the channel  $\mathcal{E}(\gamma , n)$ is not entanglement annihilating. It has been shown analytically that, on the yellow line  in  Fig.  \ref{Fig4},  the GAD channel is  entanglement annihilating  \cite{25}. In addition, it been shown numerically that, in the violet region in  Fig.  \ref{Fig4},  the GAD channel is  entanglement annihilating  \cite{30}. Note that on the yellow line and in the violet region   the GAD channel is
   not entanglement breaking.

Also note that the region in the $(\gamma, n)$-plane for which $\mathcal{E}(\gamma , n)$ is entanglement annihilating, but not entanglement breaking, does not includes the line $\gamma =1$, i.e., $\mathcal{E}(\gamma , n)$ is invertible in this region. Therefore, the GAD channel with parameters in the violet region and on the yellow line satisfies all the three required properties mentioned at the beginning of this section.
So, with the  parameters $\gamma$ and $n$ in this region, the non-positive localized map $\mathcal{E}^{-1}(\gamma , n)\otimes \mathrm{id}$ can generate entanglement.

The next question is how to implement physically the non-positive localized map $\mathcal{E}^{-1}(\gamma , n)\otimes \mathrm{id}$. A simple way for
physical implementation of a quantum channel  is  using its Stinespring
representation \cite{24}. 
For the GAD channel  $\mathcal{E}_A(\gamma , n)$, acting on the qubit $A$, we have
\begin{equation}
\label{eq:24}
\begin{aligned}
\rho_{A}^{\prime}=\mathcal{E}_A(\gamma , n) [\rho_A]  \qquad\qquad\qquad \   \\
=\mathrm{Tr}_{E_A} \circ \mathrm{Ad}_{U_{AE_A}} [\rho_{A}\otimes\tilde{\omega}_{E_A}],
\end{aligned}
\end{equation}
where    $\rho_A$ is an arbitrary initial state of the qubit $A$, the environment $E_A$ is also a qubit and
\begin{equation}
\label{eq:25}
\tilde{\omega}_{E_A}=(1-n) \vert 0_{E_A}\rangle\langle 0_{E_A}\vert +  n \vert 1_{E_A}\rangle\langle 1_{E_A}\vert,
\end{equation}
is its fixed initial state, and 
\begin{equation}
\label{eq:26}
U_{AE_A}=\left( 
\begin{matrix} 
1 & \ 0 & 0 & 0  \\
0 &  \sqrt{1-\gamma} &  i\sqrt{\gamma} & 0 \\
0 &  i\sqrt{\gamma} &  \sqrt{1-\gamma} &  0 \\
0 & \ 0 & 0 & 1 
\end{matrix}
\right), 
\end{equation} 
is the unitary time evolution of the whole system-environment $AE_A$  \cite{24}. Note that $\tilde{\omega}_{E_A}$ in Eq.  \eqref{eq:25} can be interpreted as a thermal state, where $n$ is the probability of finding the qubit 
$E_A$ in its exited state which is related to its temperature.

Therefore, for the bipartite system $S=AB$, the quantum channel  $\mathcal{E}_A(\gamma , n)\otimes \mathcal{E}_B(\gamma , n) $ can be implemented physically as
\begin{equation}
\label{eq:27}
\begin{aligned}
\rho_{AB}^{\prime}=\mathcal{E}_A(\gamma , n)\otimes \mathcal{E}_B(\gamma , n) [\rho_{AB}]  \qquad\qquad\qquad\qquad\qquad   \\
=(\mathrm{Tr}_{E_A} \circ \mathrm{Ad}_{U_{AE_A}}) \qquad\qquad\qquad\qquad\qquad\quad\qquad  \\
\qquad\qquad \otimes (\mathrm{Tr}_{E_B} \circ \mathrm{Ad}_{U_{BE_B}}) [\tilde{\omega}_{E_A}\otimes\rho_{AB}\otimes\tilde{\omega}_{E_B}],
\end{aligned}
\end{equation}
where $\rho_{AB}$ is an arbitrary initial state of the system $S=AB$, but $\tilde{\omega}_{E_A}$ and $\tilde{\omega}_{E_B}$ are fixed initial states of the local environments $E_A$ and $E_B$, respectively.
In addition, $\tilde{\omega}_{E_B}$ and $U_{BE_B}$ are defined similarly as Eqs. \eqref{eq:25}  and \eqref{eq:26}, respectively.

\begin{figure*}
\begin{center}
\includegraphics[width=15 cm]{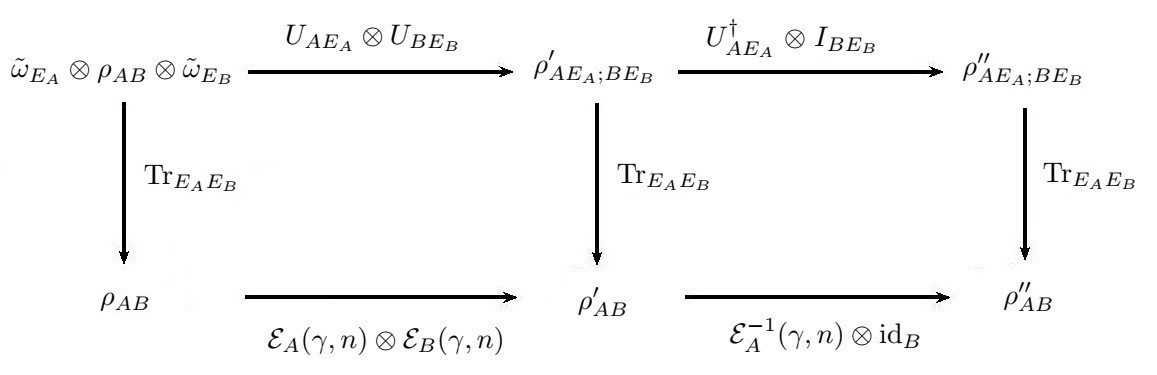}
\end{center}
\caption{One way for performing physically the non-positive localized map $\mathcal{E}_A^{-1}(\gamma , n)\otimes \mathrm{id}_B$: First, consider a factorized initial state of the system-environment as $\tilde{\omega}_{E_A}\otimes\rho_{AB}\otimes\tilde{\omega}_{E_B}$, where $\rho_{AB}$ is arbitrary, but $\tilde{\omega}_{E_A}$ and $\tilde{\omega}_{E_B}$ are fixed states as  Eq. \eqref{eq:25}. Then, perform the unitary time evolution $U_{AE_A}\otimes U_{BE_B}$, with local unitary operators $U_{AE_A}$ and $U_{BE_B}$ as Eq. \eqref{eq:26}. This results in $\rho_{AE_A;BE_B}^\prime$. Tracing over the environments $E_A$ and $E_B$ gives us the initial state of the map $\mathcal{E}_A^{-1}(\gamma , n)\otimes \mathrm{id}_B$, i.e., the secondary initial state of the system $\rho_{AB}^{\prime}$. Next, performing the unitary time evolution  ${U_{AE_A}^\dagger}\otimes I_{BE_B}$, where ${U_{AE_A}^\dagger}$ is the Hermitian conjugate of Eq. \eqref{eq:26} and $I_{BE_B}$ is the identity operator on $BE_B$, leads to $\rho_{AE_A;BE_B}^{\prime\prime}$. Now, tracing over the environments gives us  $\rho_{AB}^{\prime\prime}$, which is the final state of the non-positive map 
$\mathcal{E}_A^{-1}(\gamma , n)\otimes \mathrm{id}_B$.}
\label{Fig2}
\end{figure*}

Next, consider the (secondary) set of   possible  initial states of the system-environment   $\mathcal{S}^{\prime}=\lbrace\rho_{SE}^\prime = \rho_{AE_A;BE_B}^\prime \rbrace $, where
\begin{equation}
\label{eq:28}
\begin{aligned}
\rho_{AE_A;BE_B}^\prime
= (\mathrm{Ad}_{U_{AE_A}} \otimes \mathrm{Ad}_{U_{BE_B}}) [\tilde{\omega}_{E_A}\otimes\rho_{AB}\otimes\tilde{\omega}_{E_B}].
\end{aligned}
\end{equation}
So, we have
\begin{equation}
\label{eq:29}
\begin{aligned}
\rho_{AB}^{\prime\prime}
=(\mathrm{Tr}_{E_A} \circ \mathrm{Ad}_{U_{AE_A}^\dagger})
 \otimes \mathrm{Tr}_{E_B} [\rho_{AE_A;BE_B}^\prime] \\
 =\mathrm{id}_A \otimes\mathcal{E}_B(\gamma , n) [\rho_{AB}] \qquad\qquad\qquad\quad   \\
=\mathcal{E}_A^{-1}(\gamma , n)\otimes \mathrm{id}_B  [\rho^\prime_{AB}]. \qquad\qquad\qquad 
\end{aligned}
\end{equation}
In other words, as illustrated in Fig. \ref{Fig2}, performing the localized unitary evolution ${U_{AE_A}^\dagger}\otimes I_{BE_B}$ on the set of initial states  $\mathcal{S}^{\prime}$ provides us a way for physical implementation of the non-positive localized map $\mathcal{E}_A^{-1}(\gamma , n)\otimes \mathrm{id}_B$.

The above suggested way for implementing the   non-positive map 
$\mathcal{E}_A^{-1}(\gamma , n)\otimes \mathrm{id}_B$ can also explain  why this map is entanglement exceeding.
We can follow a similar discussion as given in the last paragraphs of Sec.~\ref{sec:D}. 
Here also the time evolution of the whole system-environment is as Eq.  \eqref{eq:4}. So, Eq. \eqref{eq:19} is  valid here too, i.e., according to the  bipartition $({AE_{A}; BE_{B}})$, we have
\begin{equation}
\label{eq:30}
\begin{aligned}
\mathcal{M}(\rho_{AB})=\mathcal{M}(\tilde{\omega}_{E_A}\otimes\rho_{AB}\otimes\tilde{\omega}_{E_B}) \qquad\qquad  \  \\
 =\mathcal{M}(\rho_{AE_{A};BE_{B}}^\prime)=\mathcal{M}(\rho_{AE_{A};BE_{B}}^{\prime\prime}).
\end{aligned}
\end{equation} 
Though the amount of  entanglement between the two distinct parts $AE_{A}$ and $BE_{B}$ remains unchanged during the time evolution, its accessible part in $\rho_{AB}$, $\rho_{AB}^{\prime}$ or $\rho_{AB}^{\prime\prime}$ can vary.
For an entanglement annihilating, but not entanglement breaking, channel $\mathcal{E}(\gamma , n)$, the state $\rho_{AB}^{\prime}$ in Eq.  \eqref{eq:27} is separable. Therefore, similar to Eq. \eqref{eq:20}, all  the amount of the entanglement existing in the whole system-environment is inaccessible for the system $S=AB$ at this moment. Performing the non-positive map $\mathcal{E}_A^{-1}(\gamma , n)\otimes \mathrm{id}_B$ on this separable state, makes some amount of this  inaccessible entanglement again accessible in $\rho_{AB}^{\prime\prime}$.

It is also worth noting that a similar method as what illustrated in  Fig. \ref{Fig2} can be used to implement physically any  local map   $\mathcal{E}^{-1}\otimes \mathrm{id}$, where the map $\mathcal{E}^{-1}$ is the inverse of a quantum channel $\mathcal{E}$. In addition, if this reduced dynamics $\mathcal{E}^{-1}\otimes \mathrm{id}$ is  entanglement generating, i.e., if it maps a separable state to an entangled one, then we  conclude that it is surely a non-positive map too.

The physical domain of a Hermitian map $\Phi_{S}$  is defined as the set $\tilde{\mathcal{S}}_S =\left\lbrace \rho_S\right\rbrace$ of all possible initial states of the system $\rho_S$ for which $\rho_S^\prime=\Phi_{S}(\rho_S)$ is a valid density operator.
A general method for physical implementing any Hermitian map $\Phi_{S}$, on its physical domain $\tilde{\mathcal{S}}_S$, has been introduced in Ref. \cite{18}. In this method, the set of possible initial states of the system-environment is chosen as (the convex hull of) the set  $\tilde{\mathcal{S}} =\left\lbrace \rho_S\otimes \Phi_{S}(\rho_S) \mid \rho_S \in \tilde{\mathcal{S}}_S \right\rbrace$, and the unitary time evolution operator of the whole system-environment is the swap operator $U_{swap}$. Therefore, for each initial state of the system $\rho_S \in \tilde{\mathcal{S}}_S$, one must first construct $\Phi_{S}(\rho_S)$ as the initial state of the environment, and then implement $U_{swap}$ on the whole system-environment, to achieve $\rho_S^\prime=\Phi_{S}(\rho_S)$ as the final state of the system.

Obviously, our method in Fig. \ref{Fig2} for implementing  $\Phi_{S}=\mathcal{E}^{-1}_A\otimes \mathrm{id}_B$, for a bipartite system $S=AB$, is much simpler than the above general method. 
In fact, the method in Fig. \ref{Fig2} can be generalized directly to include any Hermitian map   $\Phi_{S}=\mathcal{E}^{-1}_S$, where $\mathcal{E}^{-1}_S$ is the inverse of an arbitrary invertible quantum channel $\mathcal{E}_S$ on the system  $S$.

Nevertheless,  note that, in our method in Fig. \ref{Fig2}, the experimenter must be able to manipulate both the system $S=AB$ and the environments $E_A$ and $E_B$ to implement $U_{AE_A}$,  $U_{BE_B}$ and  $U_{AE_A}^\dagger $. Assuming that the experimenter has access only to the system, our proposed method to construct non-positive local maps does not work.
 In other words, we have not proposed a method to implement non-positive   maps as the \textit{dynamics} of a (closed) quantum system $S$, which presents prohibitive experimental challenges.
  But, we have only proposed a method to implement such maps as the \textit{reduced dynamics} of the system $S$ interacting with its environment, assuming that the environment is effectively modelled   as a fully controllable quantum register rather than an inaccessible dissipative bath.

\section{Summary}\label{sec:F}

In this paper, we studied the entanglement dynamics of a bipartite system $S=AB$ under local non-positive maps.
Such evolution can lead to entanglement exceeding, as   was shown in the case studied in Ref. \cite{13}. 
 In addition,  we explored how such maps can be implemented physically as the reduced dynamics of the system $S$.

We discussed two procedures for implementing local non-positive maps. First,  given in Sec.~\ref{sec:B}, is in fact (the generalization of) the case studied in Ref. \cite{13}. In this method, the evolution of the whole system-environment is localized as Eq.  \eqref{eq:4}, and the initial state of the system-environment at $t=0$ is factorized as Eq.  \eqref{eq:5}. So, the reduced dynamics of the system is given by the local CP map in Eq. \eqref{eq:6}. But, changing the initial moment from $t=0$ to some appropriate $t_i >0$, results in the map in Eq.  \eqref{eq:8}, which is local, but, in general, non-positive.
The necessary requirement for this new initial moment $t_i$ is that both $\mathcal{E}_{A}(t_i)$ and $\mathcal{E}_{B}(t_i)$ must be invertible. 

Then, in Sec.~\ref{sec:C}, we studied an example of such local non-positive reduced dynamics, and showed that  this evolution can lead to entanglement exceeding in this case too. In Sec.~\ref{sec:D}, we discussed that this entanglement exceeding is not entanglement generation, but, it is, in fact, making accessible in the system $S=AB$ some amount of the supply of the 
entanglement between the two distinct parts $AE_A$ and $BE_B$.

The second method of constructing local non-positive maps, which lead to entanglement exceeding, was given in Sec.~\ref{sec:E}. There, we showed that, for a quantum channel $\mathcal{E}$ which fulfills 
 the three requirements mentioned at the beginning of the Sec.~\ref{sec:E}, the non-positive local map $\mathcal{E}^{-1}\otimes \mathrm{id}$ results in entanglement exceeding.

As an example of this method, we studied the GAD channel $\mathcal{E}(\gamma , n)$ which satisfies the three mentioned properties for a region  in the $(\gamma, n)$-plane, and so $\mathcal{E}_A^{-1}(\gamma , n)\otimes \mathrm{id}_B$ is entanglement generating in this region.

 In addition,  we showed that how the non-positive local map $\mathcal{E}_A^{-1}(\gamma , n)\otimes \mathrm{id}_B$ can be implemented physically as the reduced dynamics of the system $S=AB$ in  Fig. \ref{Fig2}.
A similar method  can be used  for any other quantum channel $\mathcal{E}_A$, which is invertible, to implement physically the  local map $\mathcal{E}^{-1}_A\otimes \mathrm{id}_B$.
In fact, the method in Fig. \ref{Fig2} is much simpler than the general method introduced in Ref. \cite{18}.

Finally, we mentioned that this localized method of physical implementation of  $\mathcal{E}_A^{-1}(\gamma , n)\otimes \mathrm{id}_B$ in  Fig. \ref{Fig2}  clarifies that here also the role of this non-positive local map is not  generating the entanglement, but, again, it makes  accessible in the system $S=AB$ some amount of the supply of the 
entanglement in the whole system-environment $SE=AE_A;BE_B$.

%


\end{document}